# Text Writing in Air

Saira Beg [1], M. Fahad Khan [2],Faisal Baig [1]
[sairabegbs,mfahad.bs,eng.faisal.baig@gmail.com]()

COMSATS Institute of Information and Technology[1]
Federal Urdu University of Arts, Science and Technology[2],
Islamabad, Pakistan.

## Abstract

This paper presents a real time video based pointing method which allows sketching and writing of English text over air in front of mobile camera. Proposed method have two main tasks: first it track the colored finger tip in the video frames and then apply English OCR over plotted images in order to recognize the written characters. Moreover, proposed method provides a natural human-system interaction in such way that it do not require keypad, stylus, pen or glove etc for character input. For the experiments, we have developed an application using OpenCv with JAVA language. We tested the proposed method on Samsung Galaxy3 android mobile. Results show that proposed algorithm gains the average accuracy of 92.083% when tested for different shaped alphabets.[Ref: [http://learnrnd.com/news.php?id=Magnetic_3D_Bio_Printing]() Here, more than 3000 different shaped characters were used. Our proposed system is the software based approach and relevantly very simple, fast and easy. It does not require sensors or any hardware rather than camera and red tape. Moreover, proposed methodology can be applicable for all disconnected languages but having one issue that it is color sensitive in such a way that existence of any red color in the background before starting the character writing can lead to false results.

## 1.    Introduction

Object tracking is considered as an important task within the field of Computer Vision. The invention of faster computers, availability of inexpensive and good quality video cameras and demands of automated video analysis has given popularity to object tracking techniques. Generally video analysis procedure has three major steps: firstly detecting of the object, secondly tracking its movement from frame to frame and lastly analyzing the behavior of that object. For object tracking, four different issues are taken into account; selection of suitable object representation, feature selection for tracking, object detection and object tracking [1]. In real world, Object tracking algorithms are the primarily part of different applications such as: automatic surveillance, video indexing and vehicle navigation etc. [1].

[Ref: ][http://learnrnd.com/news.php?id=Swarm_Clothing:Dress_Changes_Automatically]()

Another application of object tracking is human computer interaction [1-2]. Different researchers proposed many algorithms which are categorically divided into two main approaches: image based approach [3] and Glove based approach [4]. Image based approach requires images as input in order to recognize the hand (object) movements. On the other hand, Glove based approach require specific hardware which includes special sensors etc [2]. Such applications are beneficial for disabled people.

In this paper, a real time fast video based finger tip tracking and recognizing algorithm is presented. The proposed algorithm has two major tasks: first it detects the motion of colored finger in video sequence and then applies OCR (Optical Character Reorganization) in order to recognize the plotted image. Proposed method is software based approach while in literature, almost all existing finger tracking based character recognition system require extra hard ware e.g. Light Emission Diode (LED) pen, Leap Motion controller device etc. Further more for recognition they perform comparison operation in order to recognize the input character [19-21] but for our proposed system we apply OCR for character recognition. As a result of which our computational time is much reduce than [21]. The rest of the paper is organized as follows. Sections 2 and 3 present the related work and proposed methodology, respectively. Section 4 is about results and discussions and section 5 concludes the paper.

## 2. Related Work

Automatic object tracking has many applications such as; computer vision and human-machine interaction etc [1-2]. Generally, object could be a text or person which needs to be tracked. In literature, different applications of tracking algorithm are proposed. One group of researchers are used it for translating the Sign Languages [5, 9], other used it for hand gesture recognition [6-7], another group used it for text localization and detection [8, 18], tracing full body motion of object for virtual reality [10] and finger tracking based character recognition [19-21] etc.

Bragatto et.al developed a method which automatically translates the Brazilian Sign Language from video input. For real time video processing they used multilayer perceptron Neural Network (NN) with piecewise linear approximated activation function. Such activation function is used to reduce the average complexity time of NN. Moreover they used NN at two stages; in color detection stage and for hand posture classification stage. Their results show that proposed method works well with recognition rate of 99.2 % [5]. In [http://learnrnd.com/news.php?id=Magnetic_3D_Bio_Printing]() Cooper presents a method which can handle more 3D complex Bio Printing of cells more than the generalize set. In his thesis first he developed a method which reduces the requirement of tracking by identifying the errors when classification and tracking process used. In this he used two preprocessing steps; one is used for motion and other is used for identification of hand shape. He also used the viseme representation in order to increase the lexicon size. Visema is the basic position of the mouth and face when pronouncing a phoneme and it is the visual representations of phonemes. Lastly he develops a weakly supervised learning method which used to detect signs.

[Ref: ][http://learnrnd.com/news.php?id=Swarm_Clothing:Dress_Changes_Automatically]()

Araga et.al proposed a method for hand gesture recognition system by using Jordan Recurrent Neural Network (JRNN). In their system they modeled the 5 and 9 different hand postures by a sequence of representative static images. Than it take video sequence as input and start classifying the hand postures. JRNN find the input gesture after detecting the temporal behavior of the posture sequence. Moreover, they also develop a new training method. Proposed method shows the accuracy of 99.0% for 5 different hand postures whereas it obtained accuracy of 94.3% for 9 gestures [6].In [7] Yang et.al discussed an alternative solution for the problem of matching an image sequence to a model and this problem generally occurs in hand gesture recognition. Their proposed method does not rely on skin color models and can work with bad segmentation as well. They coupled both the segmentation process with recognition using intermediate grouping process. Their results show better performance with the 5% performance loss for both models.

Neumann et.al developed a method for text localization and recognition in real world images. In their paper they used hypothesis framework which can process multiple text lines. They also use synthetic fonts in order to train the algorithm and lastly they exploit the Maximally Stable Extremal Regions (MSERs) which provides robustness to geometric and illumination conditions [8].

Moreover, in [10] Wang et.al discussed the color based motion capturing system for both indoor and outdoor environments. In their proposed method they used web camera and a color shirt in order to track the object. Their proposed method result shows the proposed method can be used for virtual reality applications.

 Jari Hannuksela et.al [19], Toshio Asano et.al [20] and Sharad Vikram et.al [21] presents finger tracking based character recognition systems. In [19] author presents motion based tracking algorithm which combines two Kalman filtering and the Expectation Maximization (EM) techniques for estimating two distinct motions; finger and camera motion.  The estimation is based on the motion features, which are effectively computed from the scene for each image frame. Their main idea is to control mobile device simply by moving a finger in front of a camera.  In [20] authors discuss a visual interface that recognizes the Japanese katakana characters in the air. For tracking the hand gesture they used the Light Emission Diode (LED) pen and a TV camera. They convert the movement of pen into direction codes. The codes are normalized to 100 data items to eliminate the effect of writing speed, in which they 46 Japanese characters are defined. For single camera they achieve 92.9% accuracy of character recognition and for multiple cameras they achieve gesture direction accuracy of 9º.

In [21] a new recognition system for gesture and character input in air is presented. For detecting the finger positions they used 3D capturing device called Leap Motion controller. In their proposed method they used Dynamic Time Warping Distance (DTW) technique for searching the similar written character from the data base. For character recognition they created a data base of 29000 recordings in which pre-written characters simples are present. Data set have two parts; one is called as candidate characters data set and other is data time series characters. Candidate characters dataset contains Upper

[Ref: ][http://learnrnd.com/news.php?id=Swarm_Clothing:Dress_Changes_Automatically]()

and Lower case characters and in which each character is replicated by 5 times. Around 100 people will participate in the instrument for a total of 26000 recordings. Furthermore, they used data time series words of 3000 recordings. In their results, they show that the time series word "new" is recognized in 1.02s with a DTW window of 0 while with larger DWT window size and recordings their proposed system takes longer time in seconds.

## 3. Proposed Methodology

The system proposed in this method consists of following steps. It tracks the motion of colored finger tip, find its coordinates and plot the coordinates. After plotting coordinates optical character reorganization (OCR) is applied on plotted image, the output is matched with trained database for OCR and the most possible match is achieved and displayed as shown in figure 1.

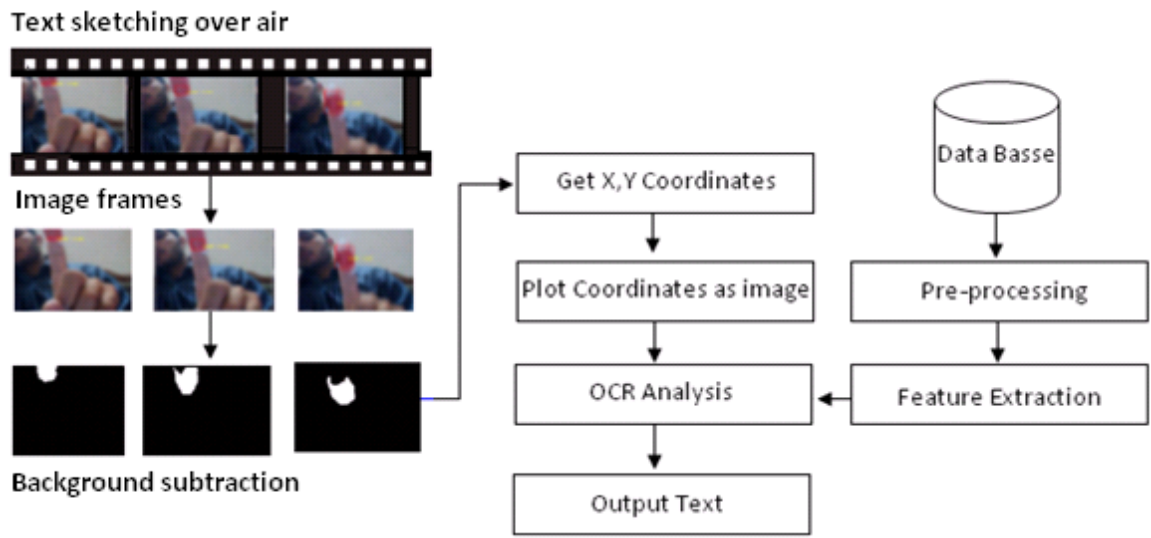

**Fig. 1:** Proposed Architecture

## 3.1 Object localization

Extracted image from video sequence is shown in figure 2. After the extraction the object is localized by the following:

**Extract color image from reference:** In this proposed method we are basically tracking the motion of the index finger which is colored red; we do not have a reference image so every previous image is the reference to next image. Now take the difference of the images and extract the color and object movement. Figure 3 shows the background abstracted image.

**Edge enhancement (EE):** The edge enhancement technique makes the object localization algorithm robust to noise, varying lighting conditions, obscuration and object fading even in the low-contrast imagery. Figure 4 shows the edge enhancement image. The edge-enhancement process consists of four operations:

## Gaussian smoothing

It is a well-known fact that the video frames captured from any camera have noise in them – at least to some extent, especially when the ambient light around the sensor is low [13, 14, 15 ]. If the frames are extracted from a compressed video clip instead of camera, they usually contain undesired artifacts in addition to noise [11, 12]. The smoothing process attenuates the sensor noise and reduces the effects of artifacts, resulting in less number of false edges in our subsequent operation (i.e. edge detection).

The average and box filters could be used to attenuate the noise and artifacts in the images, but they introduces unwanted blur resulting in the loss of the fine detail of the object. On the contrary, the Gaussian smoothing filter does the same job without sacrificing the fine detail of the object. Thus, we apply a $w \times w$ Gaussian smoothing filters with standard deviation,$\sigma_w$, on the search window and the template. As shown in equation 1. [16]

$$\sigma_w = 0.3 \left( \frac{w}{2} - 1 \right) + 0.8$$

## Edge detection

We use edge-enhanced gray-level images instead of actual gray-level images in this because edge images are less sensitive to lighting conditions. We apply the standard horizontal and vertical Sobel masks on the smoothed image, and get the two resulting images of horizontal and vertical derivative approximation which are *Eh* and *Ev*. Then, we obtain the gradient magnitude image, *E*. This image is normally obtained as in equation 2. [17]

$$E(i,j) = \sqrt{E_h^2(i,j) + E_v^2(i,j)}$$

Where $i = 1, 2... U - 1, j = 1, 2... V - 1$, where *(U, V) = (K, L)* for the template, and *(U, V) = (M, N)* for the search-window.

## Normalization

We have found out in our experiments that the dynamic range of the edge image, E, is often too narrow towards darker side as compared to the available pixel-value range [0, 255], especially in low-contrast imagery. Conventionally, the edge image is converted into a binary image using a predefined threshold; however, this approach does not work well in a template matching application, because the rich content of the gray-level edge-features of the object is lost in the process of binarization. We enhance the edges using a normalization procedure given by equation 3:

$$E_n(i,j) = \left( \frac{255}{E_{max} - E_{min}} \right) [E(i,j) - E_{min}]$$

[Ref: ]http://learnrnd.com/news.php?id=Swarm_Clothing:Dress_Changes_Automatically_

where *En* is the normalized edge image, 255 is the maximum value a pixel can have, and *Emin* and *Emax* are the minimum and maximum values in the un-normalized edge image, *E*, respectively. The normalization stage effectively tries to stretch the histogram of the image in the whole range [0, 255]. As a result, the contrast between the object and the background is also enhanced. This contrast increases the difference between the background and the object.

**Thresh holding**

Never the less, in order to remain in the safe side and quench the false edges due to smoothed noise and artifacts, a thresholding operation is performed as:

$$E_{nt}(i,j) = \begin{cases} E_n(i,j) & if\ E_n(i,j) \geq 50 \\ 0 & otherwise \end{cases}$$

Where $E_{nt}$ is the normalized and threshold edge image. It may be noted that $E_{nt}$ is not a binary image, but an edge-enhanced gray-level image adequately containing the important features of the object.

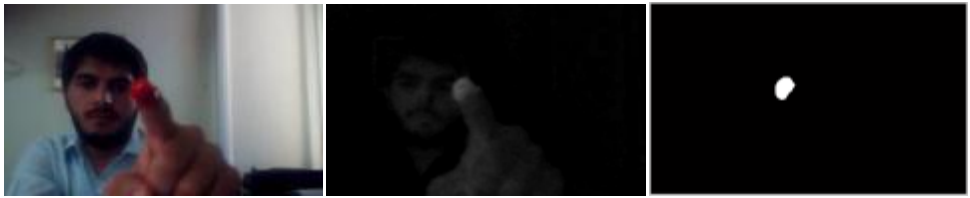

**Fig.2:** Extracted image          **Fig.3:** Background subtraction **Fig.4:** Edge enhancement

### 3.1.1   Blob Analysis

In blob analysis we extract image properties after edge enhancement and extraction of color region. We cannot directly apply blob analysis on binary images to extract the features, so to do so we first label the image. Labeling procedure is used to find the connected points in binary images which than further utilized for blob analysis. Figure 5 represents the labeled image with the connected points

[Ref: ][http://learnrnd.com/news.php?id=Swarm_Clothing:Dress_Changes_Automatically](http://learnrnd.com/news.php?id=Swarm_Clothing:Dress_Changes_Automatically)

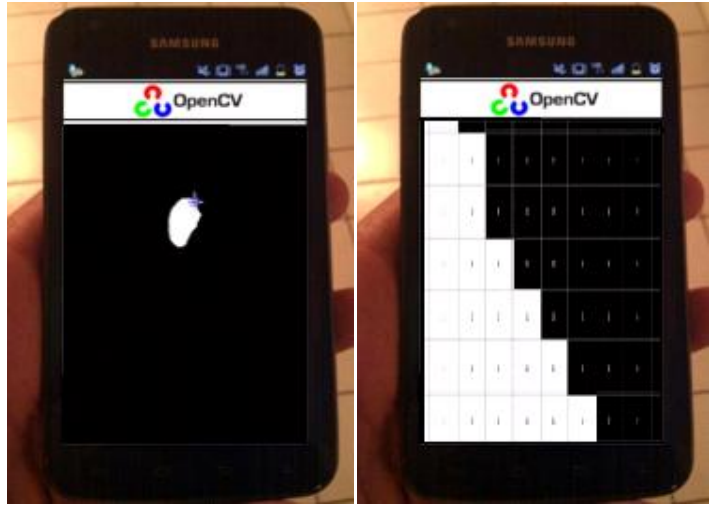

**Fig. 5:** Representing the connected points of labeled image

In blob analysis we measure the set of properties from labeled image. Positive integer elements of labeled image correspond to different regions. For example, the set of elements of labeled equal to 1 corresponds to region 1; the set of elements of Labeled equal to 2 corresponds to region 2; and so on. The set of properties are returned in an array which are Area, Bounding box, Centriod, Orientation.

### 3.1.2   Plot X, Y coordinates

After extracting the properties of object from still video frame, we track the object throughout the video frame and get its *x, y* coordinates as shown in figure 6. After extraction of *x, y* coordinates we represent these coordinates in form of image which is further take in to consideration. Here, we get the mirror images, so, we just shuffle the x coordinates rather than inverting the image while y coordinates remain same    Figure 7 shows the plotted image.

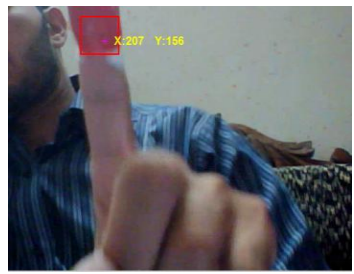    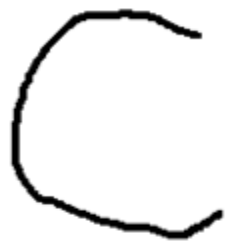

**Fig. 6:** (X, Y) (207,186) coordinates          **Fig. 7:** Plotted image

## 3.2     OCR Analysis

OCR which is abbreviation of Optical Character Recognition is a designed tool in image processing used to read and recognize the character by reading an image. The purpose of using OCR in this is, to read the resultant character after tracking the object. The OCR in this work is defines as follows:

### Load Template

A template is basically a data set used to compare the resultant image and find the relation with a text. The data set basically contains the test images that can possible be drawn while drawing a text in air.

### Convert test image in binary

To process an image it is necessary to convert an image in to binary because a binary image contains the value of pixel in 0 or 1 which are easy to deal with so. After this we label, labeled image correspond to different regions. The set of elements of labeled equal to 1 corresponds to region 1; Elements Labeled equal to 2 corresponds to region 2 and so on.

### Find connected line segment and Correlation

The purpose of finding connected line segment is to read the shape of resultant image. Figure 8 presents the resultant image.  This is done by reading labeled image line wise and store the result in an array. The test image which is shown below in figure is show the working of connected line segment.

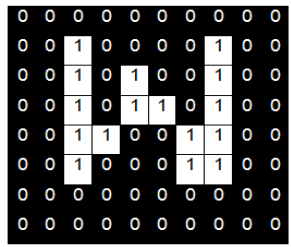

**Fig. 8:** Read the shape of resultant image

### Proposed Algorithm

Figure 9 discussed the flow chart of the proposed algorithm. in proposed algorithm we first acquire the video, extract color from video frame images, apply edge detection and enhancement to detect the object, after this apply blob analysis to bound the object and finding its *X, Y* coordinates in video frames images. After this plot these coordinates and apply optical character reorganization (OCR) analysis and take output.

[Ref: ]<u>http://learnrnd.com/news.php?id=Swarm_Clothing:Dress_Changes_Automatically_</u>

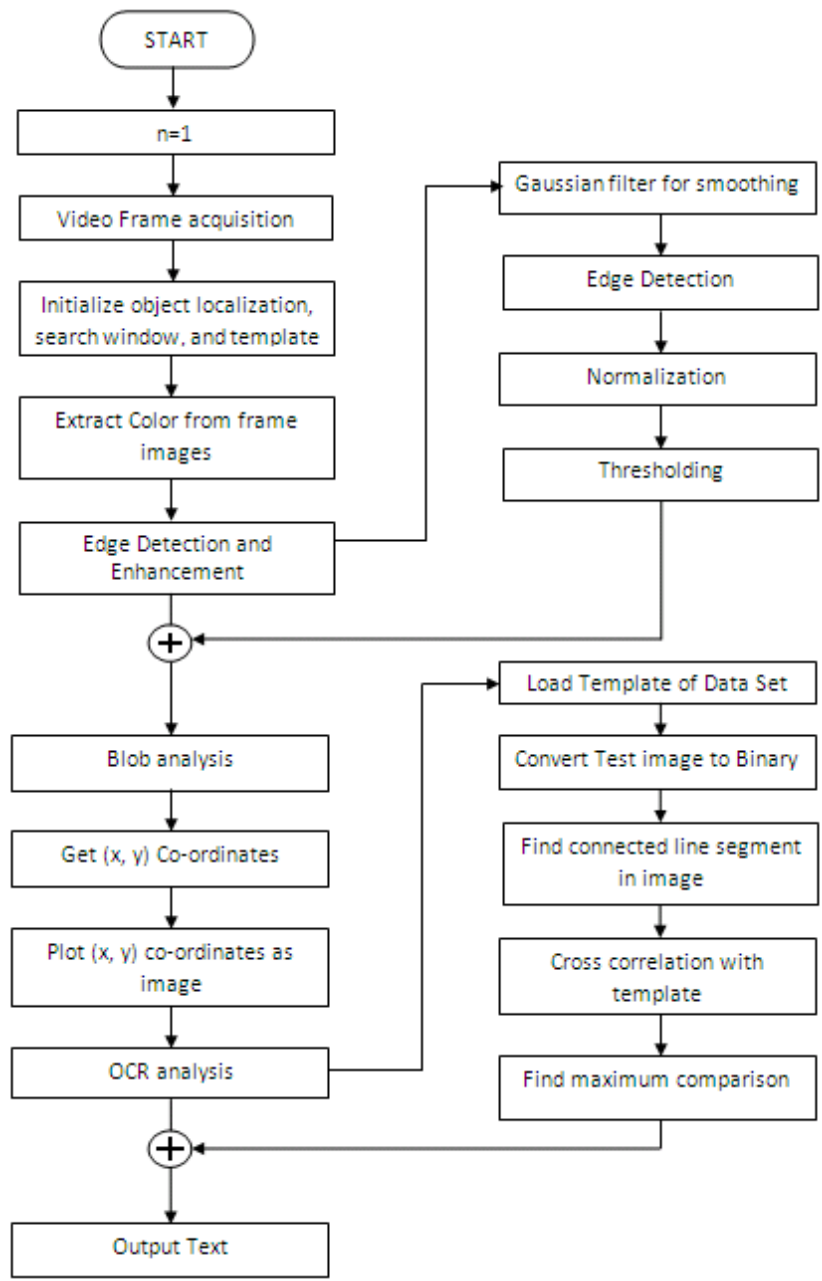

**Fig. 9:** Proposed Algorithm

## 4.       Results and Discussion

## 4.1 'W' results

[Ref: ]http://learnrnd.com/news.php?id=Swarm_Clothing:Dress_Changes_Automatically_

For experimentation, we use Samsung Galaxy3 mobile phone having Android 4.0 (Ice Cream Sandwich) Operating System. We use its frontal camera of 1.9 Mega pixels with frame rate of 30fps. We text out proposed method for both indoor and outdoor locations and found that it works well.

The system proposed in this method is track the colored finger in a video file by capturing video from any video capturing device. Then it read video file and process video frame wise by reading images; after reading frame wise images from video file, the designed algorithm extract the red color from frames. As shown in figure 10.

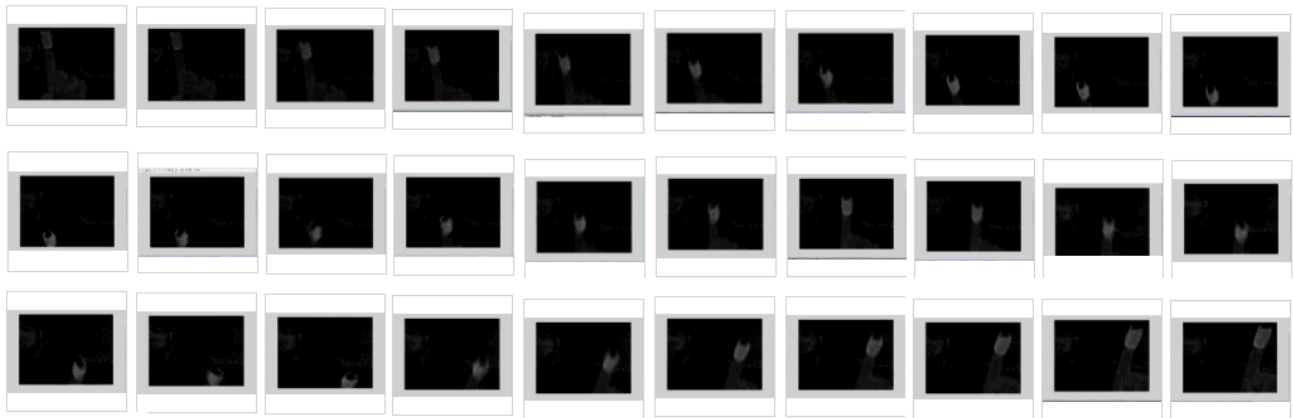

**Fig. 10:** Red Color image

Then apply edge enhancement to detect the edge of colored finger the enhanced edge is shown in figure 11.

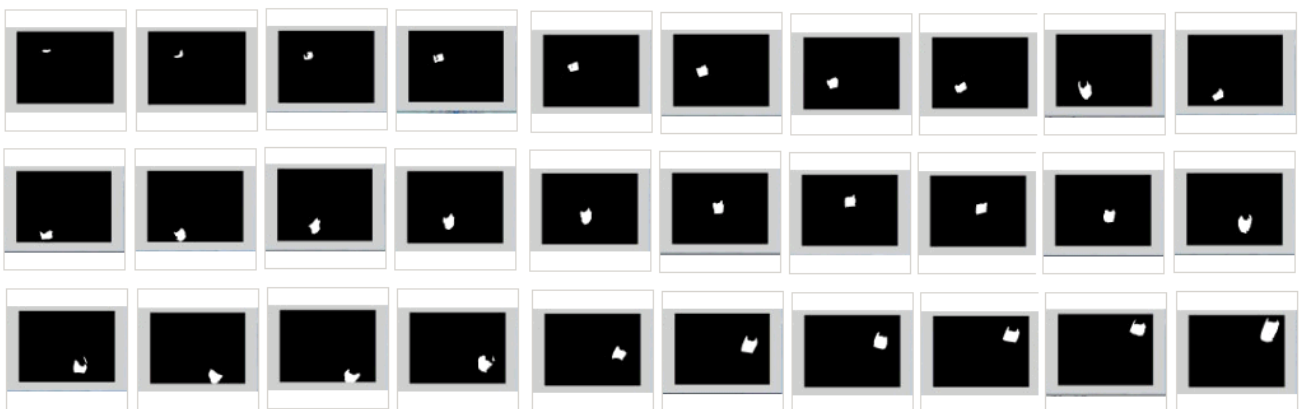

**Fig. 11:** Enhanced Edge

[Ref: ][http://learnrnd.com/news.php?id=Swarm_Clothing:Dress_Changes_Automatically_]()

After edge enhancement the designed system extract the (X,Y) coordinates of colored image by applying bounding box on the extract image, the bounding box on colored image is shown in figure 12.

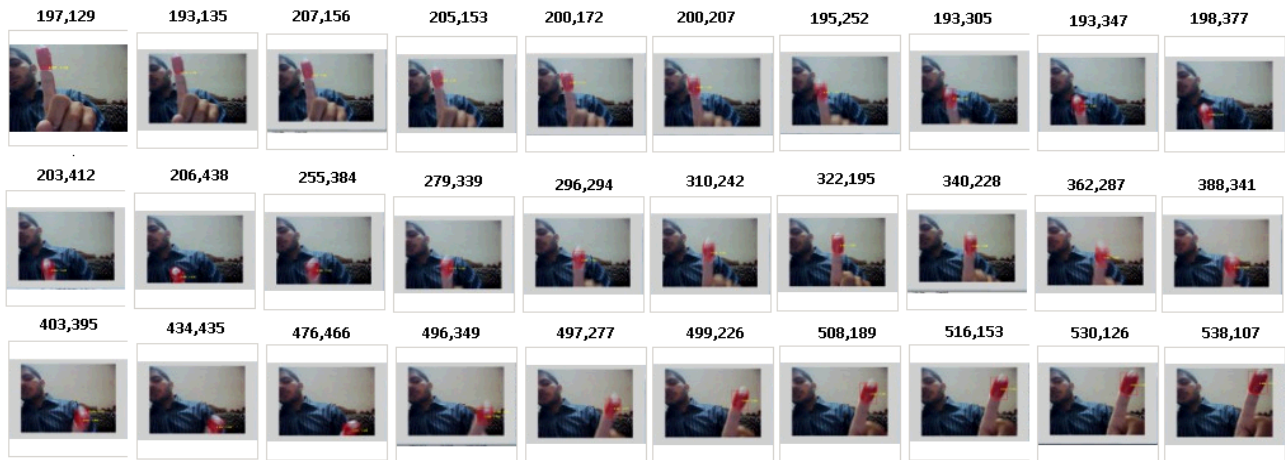

**Fig. 12:** Bounding Box with X,Y coordinates

After extraction of (X,Y) coordinates, these values are stored in an array and displayed in image form as in figure 13.

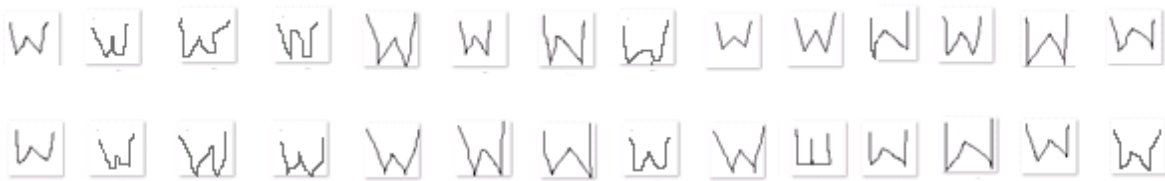

**Fig. 13:** Training set for W

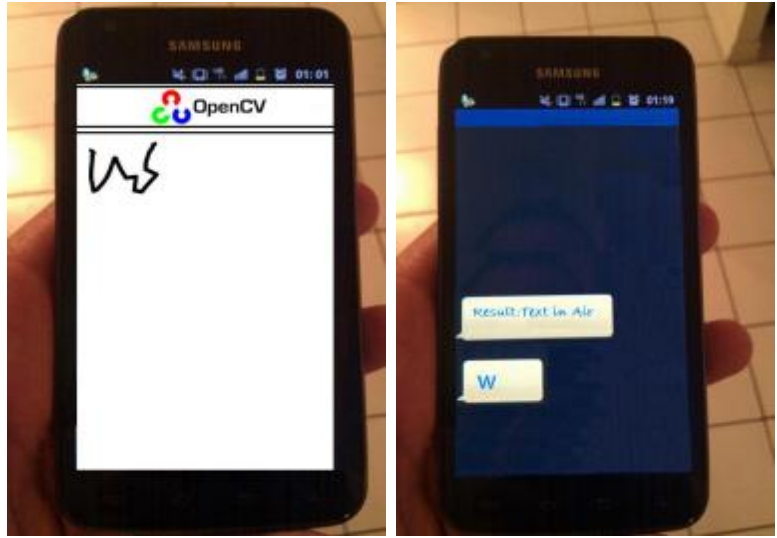

**Fig. 14:** Plotted Image          **Fig. 15:** Proposed System Result

Then this image is given to optical character recognition (OCR) module for recognition. At first it load the training test templates, in figure 14 the training set for word "W" is shown. After loading complete template for alphabetical and numeric characters, it is compared with the resultant image to find the maximum comparison and result is displayed in a text file showing resultant character as shown in figure 15.

## 4.2 "Hello World" Results

As mentioned above that our proposed method is working on color segment of the frame. And it tracks the entire region for it. Two characters could be differentiated by two means;

Use tape only for the frontal portion of the index finger and after writing each character, just twist the finger (where no tape exist) and then write the next word. When proposed method does not found colored object then it assumed that one character is completed.

Another way is to add waiting time after writing each character. The waiting time would be of some mseconds which gives duplicate frames having same features. As we discussed above that we use previous frame as a reference image and we take difference between two images so, smaller the difference value mostly near to zero in case of duplicated frame (coordinates of x, y) means that one character is complete. Both cases are clearly shown in figure 17, 18, 19

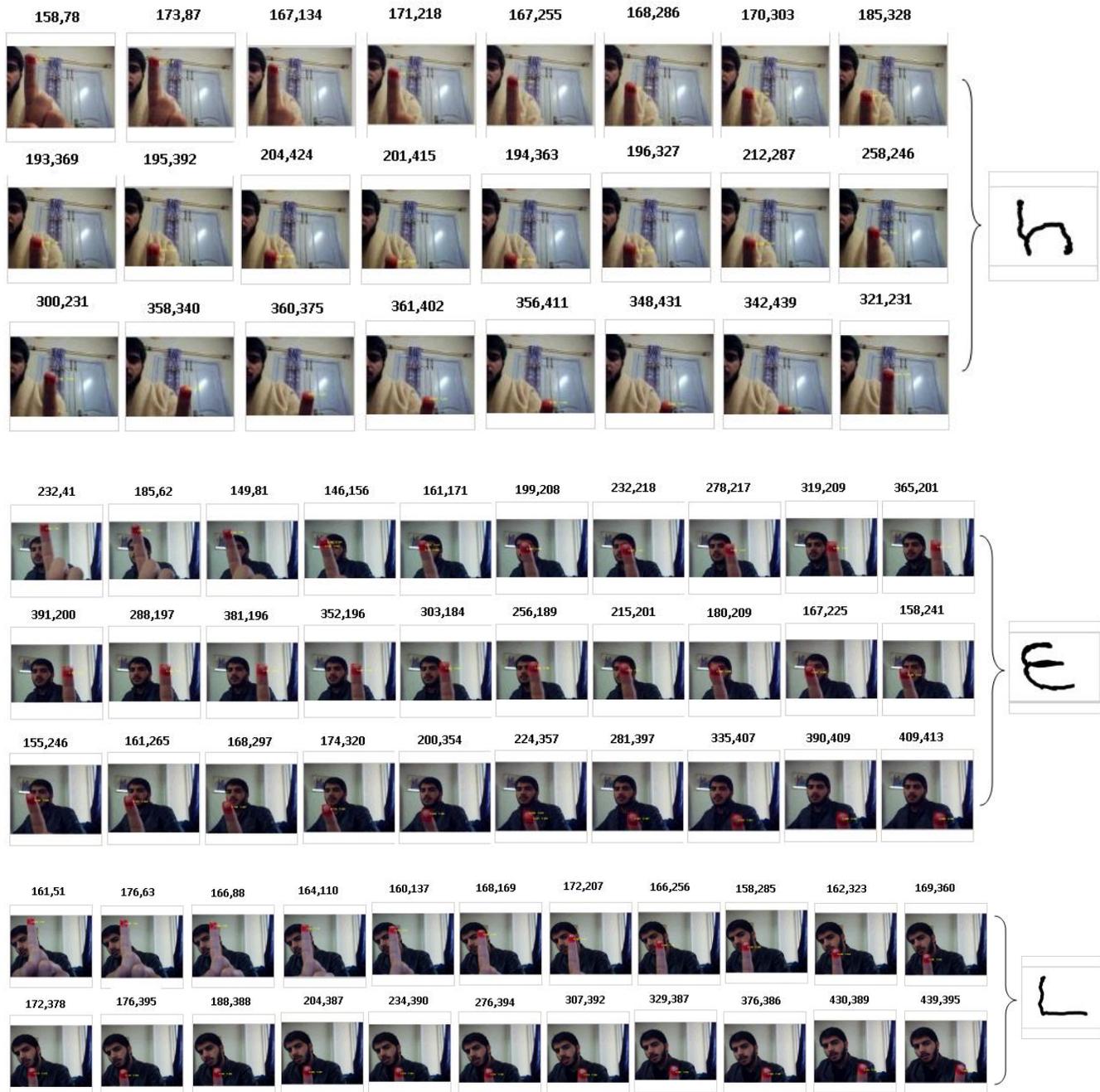

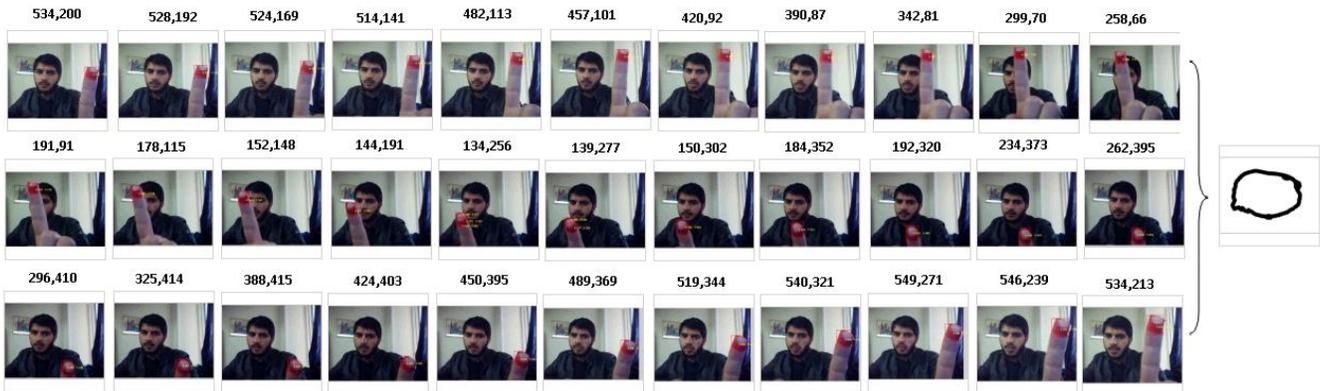

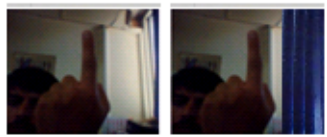 Space Bar : Back side of the finger which is non colored 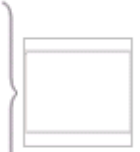

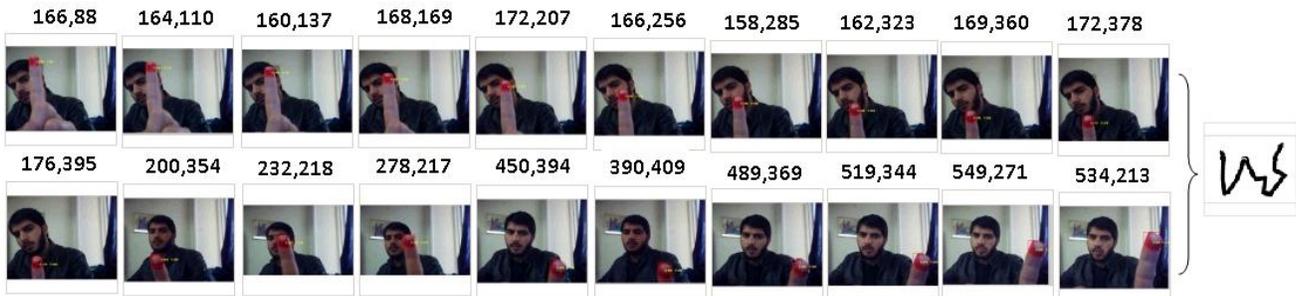

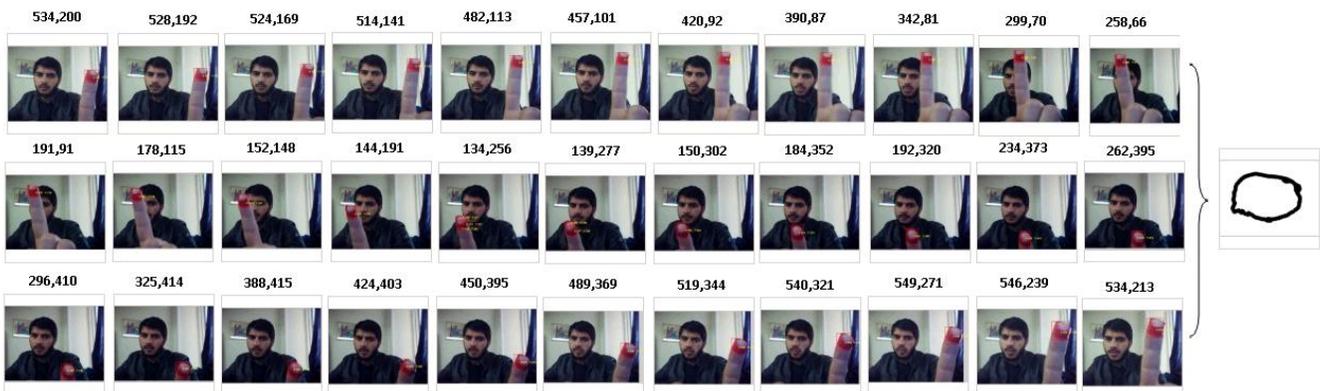

[Ref: ]http://learnrnd.com/news.php?id=Swarm_Clothing:Dress_Changes_Automatically_

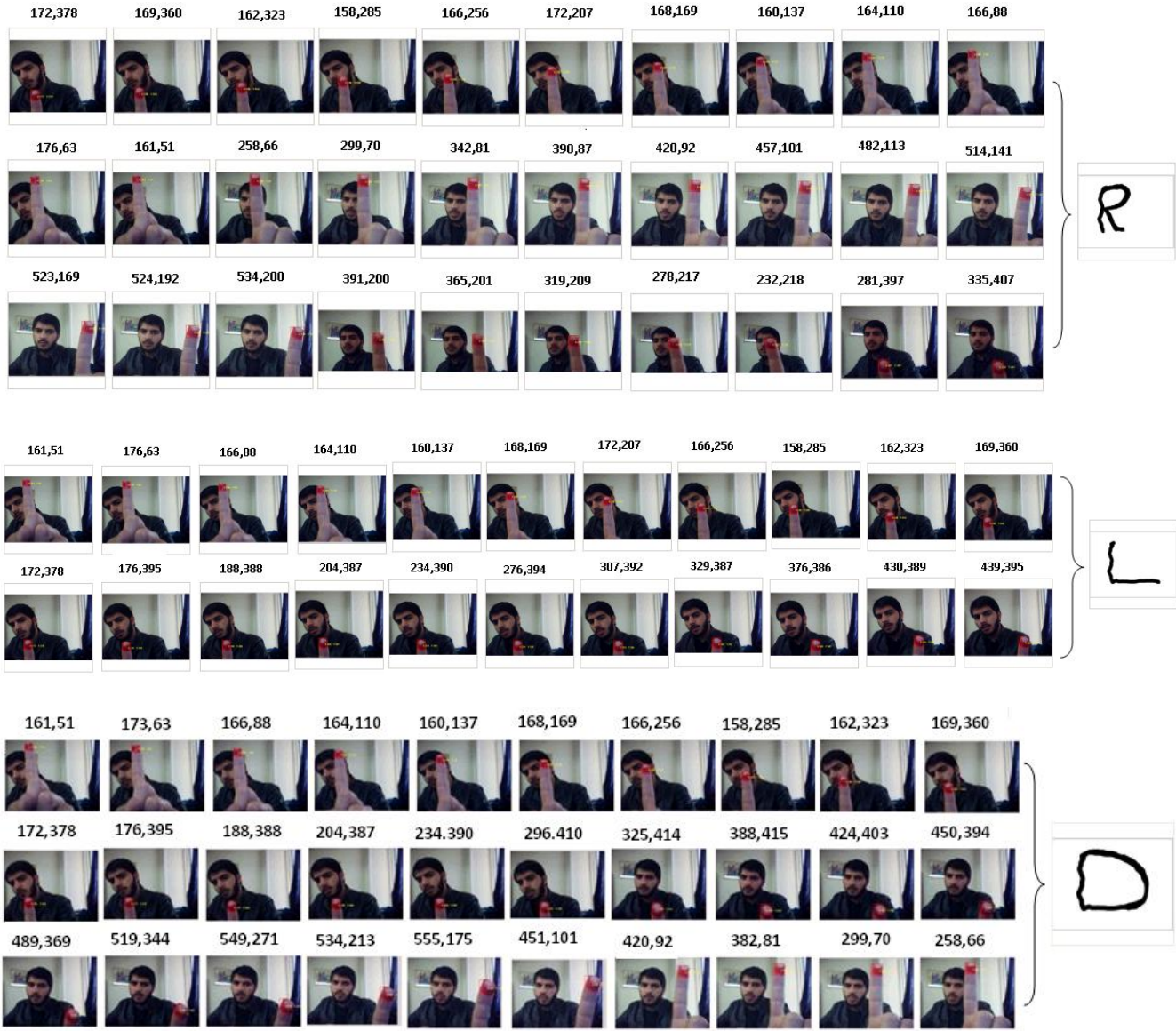

The system which is developed is trained on the following images that are shown in figure 16.

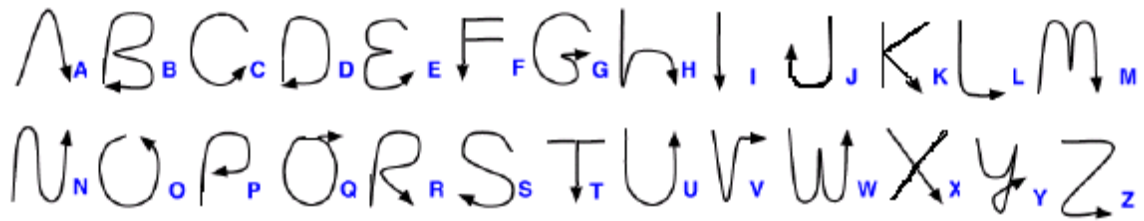

**Fig. 16:** Training images

[Ref: ]http://learnrnd.com/news.php?id=Swarm_Clothing:Dress_Changes_Automatically_

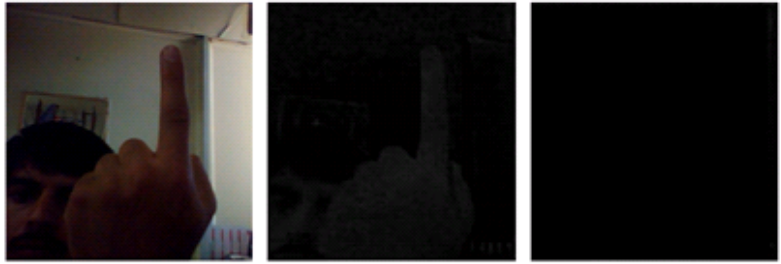

Figure 17: Back side of finger

Figure 18: Extracted color image

Figure 19: Edge detection

Furthermore, we test our proposed method for different shapes characters which were more than up to 3000. The average accuracy of the proposed method is 92.083%. Table 1 shows the proposed algorithm accuracy for each alphabet. For alphabet D and V our proposed algorithm shows least accuracy rate of 81%.

**Table1:** Accuracy rate for each alphabet

| Sr. | Alphabets | Accuracy % | Sr. | Alphabets | Accuracy % |
|---|---|---|---|---|---|
| 1 | A | 96 | 14 | N | 94 |
| 2 | B | 90 | 15 | O | 83 |
| 3 | C | 89 | 16 | P | 86 |
| 4 | D | 81 | 17 | Q | 85 |
| 5 | E | 98 | 18 | R | 90 |
| 6 | F | 100 | 19 | S | 96 |
| 7 | G | 88 | 20 | T | 98 |
| 8 | H | 98 | 21 | U | 83 |
| 9 | I | 100 | 22 | V | 81 |
| 10 | J | 100 | 23 | W | 96 |
| 11 | K | 92 | 24 | X | 92 |
| 12 | L | 93 | 25 | Y | 94 |
| 13 | M | 96 | 26 | Z | 98 |
| | | | 27 | Space | 88 |
| | | | | **AVERAGE** | **Accuracy 92.083** |

The reason behind such a least accuracy rate is that the character 'D' some time captured in such a way that it look similar to character 'O' while writing in air because writer could not see the trajectories, so the letter loci are very bad. Similar case is with the character 'V' which some time captured as character 'U and W'. Finally we calculate the reorganization speed of the proposed system. Table 2 present the recognition of different characters after getting the X, Y coordinates. Results shows that our proposed is fast than the existing method present in [21]. In their paper they showed that for word "new" their proposed method require 1.02 seconds where with the same number of words our method require 0.663678 seconds.

[Ref: ]http://learnrnd.com/news.php?id=Swarm_Clothing:Dress_Changes_Automatically

Table 2: Time require for recognition

| Number of Letters | Time elapsed by proposed method in sec |
|---|---|
| 1 | 0.600120 |
| 2 | 0.640000 |
| 3 | 0.663678 |
| 4 | 0.681000 |
| 5 | 0.690130 |
| 6 | 0.716600 |
| 7 | 0.745789 |
| 8 | 0.768872 |
| 9 | 0.769687 |
| 10 | 0.810676 |
| 11 | 0.827301 |
| 12 | 0.856799 |
| 13 | 0.870148 |
| 14 | 0.888103 |
| 15 | 0.887607 |

## 5.      Conclusion

This paper presents a video based pointing method which allows writing of English text in air using mobile camera. Proposed method have two main tasks: first it track the colored finger tip in the video frames and then apply English OCR over plotted images in order to recognize the written characters. Moreover, proposed method provides a natural human-system interaction in such way that it do not require Keypad, Pen or Glove etc for character input. It just requires a mobile camera and red color for reorganization a finger tip. For the experiments, we have developed an application using OpenCv with JAVA language. The proposed method gains the average accuracy of 92.083% in order to recognize the accurate alphabets. The overall writing delay gained by the proposed method was 50 ms per character. Moreover, proposed methodology can be applicable for all disconnected languages but having one serious issue that it is color sensitive in such a way that existence of any red color in the background before starting the analysis can lead to false results.

[Ref: ][http://learnrnd.com/news.php?id=Swarm_Clothing:Dress_Changes_Automatically]()